\documentclass{iopart}
\usepackage{graphicx}

\begin{document}

\title{Open heavy flavor measurements with the PHENIX experiment at RHIC}

\author{Ralf Averbeck\dag for the PHENIX Collaboration}

\address{\dag\ Department of Physics and Astronomy,
Stony Brook University,\\
Stony Brook, New York 11794-3800,
USA}

\ead{Ralf.Averbeck@stonybrook.edu}

\begin{abstract}
The PHENIX experiment has measured single electron spectra at RHIC in 
proton-proton (p-p), deuteron-gold (d-Au), and gold-gold (Au-Au) collisions 
at an available energy per nucleon-nucleon pair of $\sqrt{s_{NN}}$ = 200 GeV. 
Contributions from photonic sources, {\it i.e.} Dalitz decays of 
light mesons and photon conversions, are subtracted from the inclusive
spectra. The remaining non-photonic single electron spectra are dominated 
by semileptonic decays of particles carrying heavy flavor. Implications 
of these systematic measurements for heavy flavor production in cold and 
hot nuclear systems are discussed.
\end{abstract}

\section{Introduction}
Particles carrying heavy flavor, {\it i.e.} charm or beauty quarks, are 
sensitive probes of the hot and dense medium created in high energy
heavy-ion collisions. 

The production of heavy flavor quark-antiquark pairs proceeds mainly via 
gluon-gluon fusion in the very early stage of the collision, not only 
through leading order pair production processes but also via higher order 
QCD mechanisms involving flavor excitation or gluon splitting diagrams
\cite{Nor00}. While the total yield is sensitive to the initial gluon 
density \cite{App92,Mue92} as well as to nuclear effects such as shadowing, 
the phase space distribution of heavy flavor should shed some light on the 
dominant production mechanisms in inital state hard scattering processes. 
In addition, thermal production from gluon-gluon scattering in later stages 
of the collision might play a role, providing some sensitivity on the initial 
temperature reached in a nuclear collision.
 
Once being produced, a heavy quark-antiquark pair propagates through the
collision zone and forms either a bound quarkonium state or separates and 
hadronizes into two particles carrying open heavy flavor. In case a deconfined 
phase of nuclear matter is formed during the collision, the yield of heavy 
quarkonia might either be reduced due to the expected screening of the QCD 
attractive potential \cite{Mat86} or possibly even enhanced due to 
quark-antiquark coalescence \cite{The01} or statistical recombination 
\cite{And03}. 
While open heavy flavor measurements will provide an essential baseline for 
quarkonia suppression/enhancement studies \cite{dongjo}, they are of prime 
interest in their own right as well. Modifications of the gluon distribution 
function in cold nuclear matter should manifest themselves in differences in 
the heavy flavor yields and spectra between p-p and p-A collisions.
High $p_T$ hadrons carrying light quarks only are strongly suppressed
at central rapidity in Au-Au collisions at RHIC 
\cite{highpt1,highpt2,highpt3,highpt4}.
Medium induced gluon radiation, which might be the mechanism driving the
observed suppression, could be reduced for heavy quarks 
\cite{dead_cone1,dead_cone2}.
Therefore, a comparison of the spectra of particles carrying heavy flavor, 
or their decay products, produced in elementary and nuclear collision
systems should reveal crucial information regarding the mechanisms responsible
for the observed energy loss of hard scattered partons in heavy-ion collisions
at RHIC. Further insight regarding the interaction of heavy quarks with the
nuclear medium can be gained from the measurement of the elliptic flow
strength $v_2$, which addresses the question whether collectivity is bourne
out already on the parton level at RHIC \cite{batsouli,greco}.

\section{Experimental techniques}
The direct reconstruction of heavy-flavor decays, {\it e.g.} 
$D^0 \rightarrow K^- \pi^+$, is difficult in the high-multiplicity 
environment of a heavy-ion collision.
An alternative approach is to determine the contributions from semileptonic 
heavy-flavor decays, {\it e.g.} $D \rightarrow e K \nu$, to single lepton 
and lepton pair spectra.

At RHIC, the PHENIX experiment \cite{phenix_nim} is ideally suited for such 
studies since it is optimized for the measurement of leptons.
Electrons are measured in the PHENIX central arm spectrometers, which 
cover the pseudorapidity range $|\eta| < 0.35$.
At forward and backward pseudorapidity ($1.2 < |\eta| < 2.4$) two
dedicated muon spectrometers are available.

Up to now, the PHENIX open heavy-flavor program focuses on the analysis
of single electron spectra, $(e^+ + e^-)/2$, measured at central rapidity.
The momenta of charged particles are determined via tracking through a 
magnetic field using drift and pad chambers. 
Electron candidates are identified by associated hits in a ring imaging 
Cerenkov detector (RICH) passing a ring shape cut, and they are confirmed
by an associated shower in an electromagnetic calorimeter (EMC) with an energy 
that is close to the measured momentum.
Accidental coincidences between RICH hits and hadron tracks lead to a
background level of about 10\%.  
Employing an event-mixing technique, this background is estimated and
statistically subtracted from the electron sample.

The sources contributing to the inclusive electron spectra can be grouped
into two classes:

\begin{itemize}
\item {\bf Photonic} electrons originate from Dalitz decays of light 
                     neutral mesons ($\pi^0$, $\eta$, $\rho$, $\omega$,
		     $\eta'$, and $\phi$) and conversions of photons
		     in material. 
\item {\bf Non-photonic} electrons are primarily from semileptonic decays
                         of heavy flavored particles. On the level of a
			 few percent, dielectron decays of light vector 
			 mesons ($\rho$, $\omega$, and $\phi$) and weak 
			 decays of kaons contribute to the non-photonic
			 electron spectra as well, in particular at
			 low $p_T$.
\end{itemize}

The non-photonic electron spectra are constructed by measuring the inclusive
electron spectra and then subtracting the photonic contributions. The
photonic electron spectra are determined mainly via two methods:

\begin{itemize}
\item For the {\bf cocktail} method the contributions from Dalitz (and
      dielectron) decays are calculated with a hadron decay generator
      \cite{phenix_au130}, where the hadron yields and phase space
      distributions are parameterized according to available data.
      The electron spectra from photon conversions are very similar in
      shape to the corresponding Dalitz electron spectra, as confirmed 
      in detailed simulations. The yield ratio depends on the material
      in the detector acceptance and is typically in the order of one.
\item For the {\bf converter} method a thin photon converter (1.7 \%
      $X_0$ brass) is introduced into the detector acceptance close 
      to the interaction point for a fraction of the running period.
      By measuring the electron spectra with and without the converter
      being present in the setup, the fraction of electrons originating
      from photonic sources can be inferred \cite{phenix_au200}.
\end{itemize}

These two techniques are complementary to each other. The converter
method allows for a precise measurement of the non-photonic electron
spectrum at low $p_T$ (in particular below 1~GeV/c), where the ratio 
of non-photonic to photonic electrons is so small ($\le$20\%) that 
the cocktail method suffers from substantial systematic uncertainties, 
mainly due to the independent normalization of inclusive electron spectra 
and the cocktail input. Therefore, the converter method is the tool of 
choice for the measurement of the total non-photonic electron cross section, 
since this is dominated by the electron yield at low $p_T$.
Towards higher $p_T$, the ratio of non-photonic to photonic electrons
increases rapidly, eliminating the main uncertainty in the cocktail 
subtraction technique. 
At the same time, the converter method suffers from statistics limitations
due to the decreasing photonic electron yield with increasing $p_T$.
Hence, for the measurement of the spectral shapes of non-photonic electrons
it is beneficial to use the cocktail method.
A third technique is the so-called $e\gamma$-coincidence technique, which
relies on the fact that electrons from photonic sources are accompanied
by a photon. 
This method can provide a further independent cross check of the photonic
electron measurement, but the requirement of registering a photon in
coincidence with an electron results in a significantly reduced statistical
precision compared with the other two techniques.

\section{The reference: p-p collisions at $\sqrt{s}$ = 200 GeV}

\begin{figure}
\begin{center}
\includegraphics[width=0.49\textwidth]{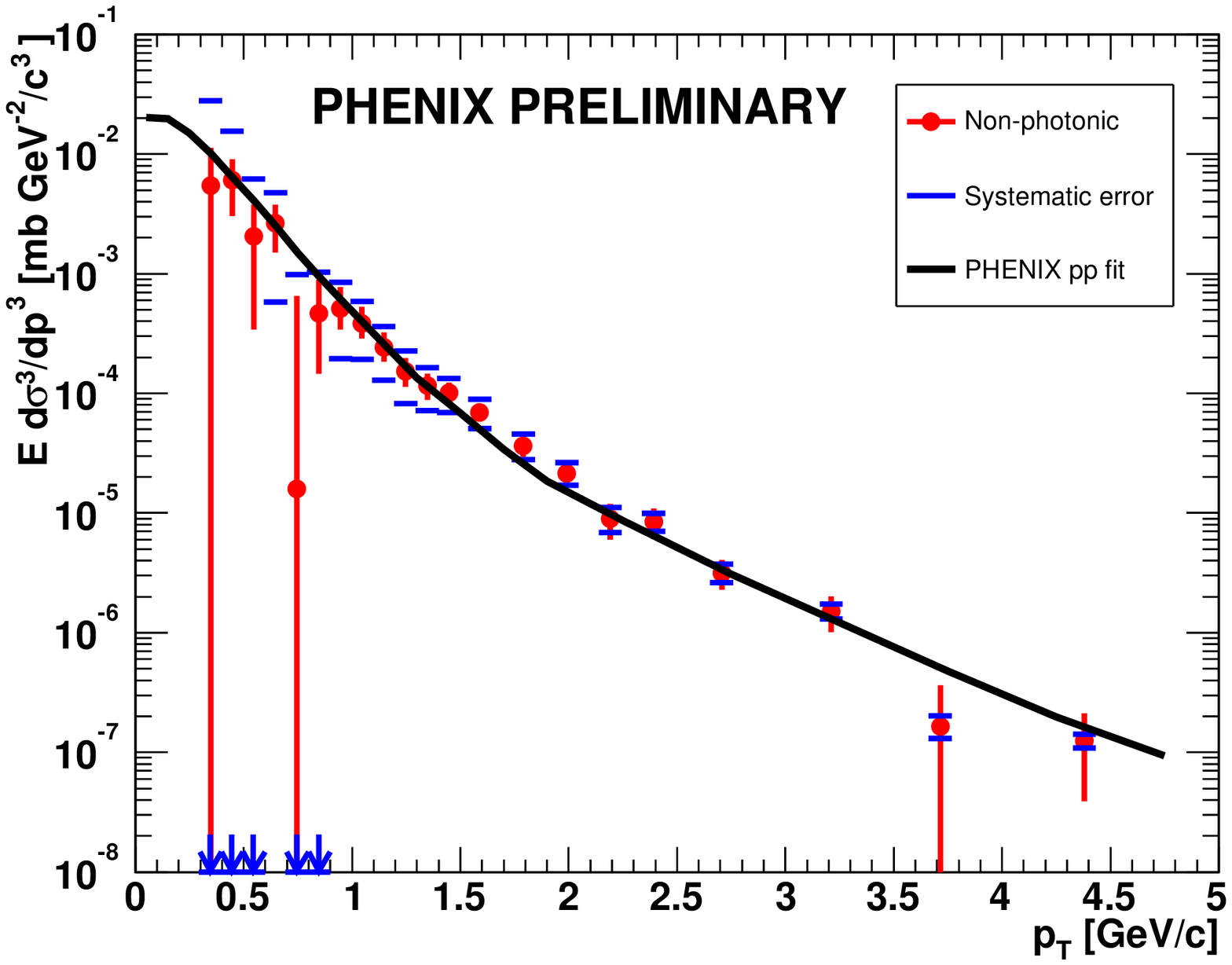}
\includegraphics[width=0.49\textwidth]{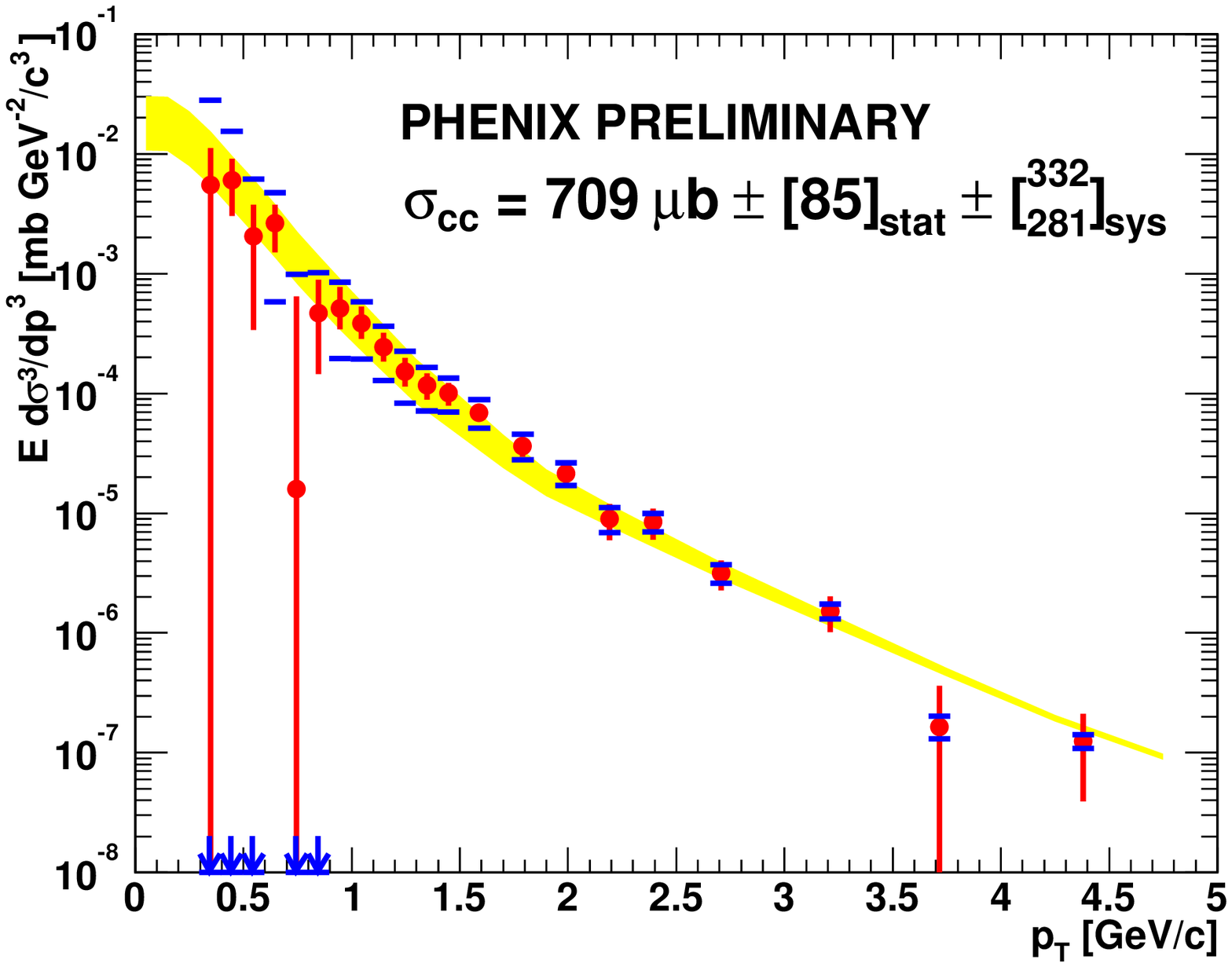}
\caption{Invariant spectra of electrons from non-photonic sources measured 
in p-p collisions at $\sqrt{s}$ = 200 GeV together with an empirical fit 
(left panel) and the error band used for estimating the charm production 
cross section (right panel), as discussed in Ref.~\cite{kelly}. At low $p_T$, 
the systematic errors are substantial due to the small ratio of the 
non-photonic signal to the inclusive electron yield.}
\label{fig1}
\end{center}
\end{figure}

A baseline for open charm production in nuclear collisions is given by 
the preliminary spectrum of non-photonic electrons from p-p collisions 
at $\sqrt{s}$ = 200 GeV shown in Fig.~\ref{fig1} \cite{kelly}. The spectrum 
was derived from a cocktail analysis and confirmed by converter and 
$e\gamma$ analyses. The analyzed RHIC Run-2 data sample comprises 15M 
minimum-bias triggers and 420M sampled minimum-bias triggers associated 
with an ERT trigger that required a coincidence between calorimeter and RICH.
It is interesting to note that for $p_T > 1.5$~GeV/c the electron spectrum
is harder than expected from PYTHIA calculations of charm and bottom pair 
production, where the PYTHIA parameters were tuned to existing data from 
lower energy experiments \cite{phenix_au130}. 
This might be due to the fragmentation function being harder than implemented 
in default PYTHIA or it could point to the fact that other production 
mechanisms beyond leading order gluon-gluon and quark-antiquark fusion play 
an important role at RHIC energies. 
Since in the low $p_T$ region, which dominates the total cross section, the
agreement between PYTHIA and the data is quite reasonable, the total charm
production cross section was determined by extrapolating with PYTHIA from 
the measured  central rapidity region to the full phase space.
The cross section obtained via this extrapolation technique is 
$\sigma_{c\bar{c}} = 709 \mu b \pm [85]_{stat} \pm [^{332}_{281}]_{sys}$.

\section{Cold nuclear matter: d-Au collisions at $\sqrt{s_{NN}}$ = 200 GeV}

\begin{figure}
\begin{center}
\includegraphics[width=1.0\textwidth]{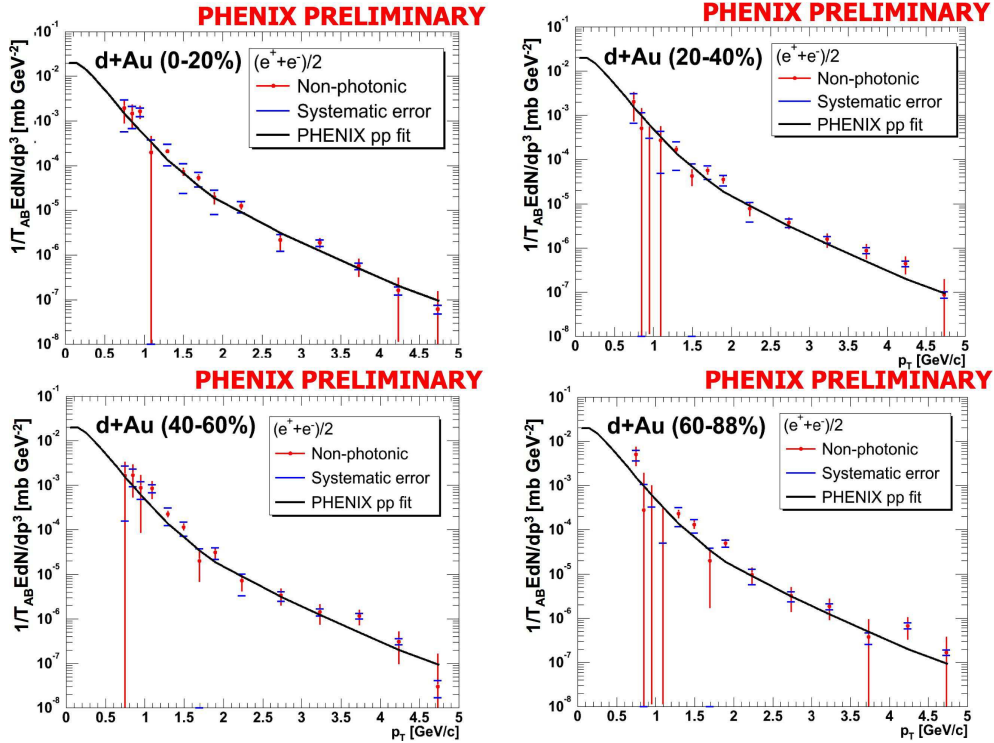}
\caption{Invariant spectra of electrons from non-photonic sources
measured in d-Au collisions at $\sqrt{s_{NN}}$ = 200 GeV scaled down by
the nuclear thickness function $T_{AB}$ in four centrality classes 
\cite{kelly}. The solid line is the best fit to p-p data as shown in
Fig.~\ref{fig1}.}
\label{fig2}
\end{center}
\end{figure}

The measurement of non-photonic electron spectra from d-Au collisions
in RHIC Run-3 allows to investigate whether any cold nuclear matter
effects, {\it e.g.} different gluon distribution functions for protons
and gold nuclei, modify charm and/or bottom production.
Fig.~\ref{fig2} shows preliminary non-photonic electron spectra from 
d-Au collisions at $\sqrt{s_{NN}}$ = 200 GeV in four centrality classes
\cite{kelly}, scaled down by the nuclear thickness function $T_{AB}$
\cite{tab200_dAu}, in comparison with the reference fit to p-p data
from Fig.~\ref{fig1}. 
The preliminary d-Au analysis uses the converter subtraction technique,
based on a converter-in data set of 5M minimum-bias events plus 312M
minimum-bias equivalent events sampled by the ERT trigger and a 
converter-out data set of 5M minimum-bias events plus 600M ERT sampled 
events, respectively.
Within the experimental uncertainties the d-Au non-photonic electron
spectra are in good agreement with the p-p spectra for all centrality
classes, indicating no strong cold nuclear matter effects on heavy
flavor production at central rapidity at RHIC.

\section{Hot nuclear matter: Au-Au collisions at $\sqrt{s_{NN}}$ = 200 GeV}

\begin{figure}
\begin{center}
\includegraphics[width=0.7\textwidth]{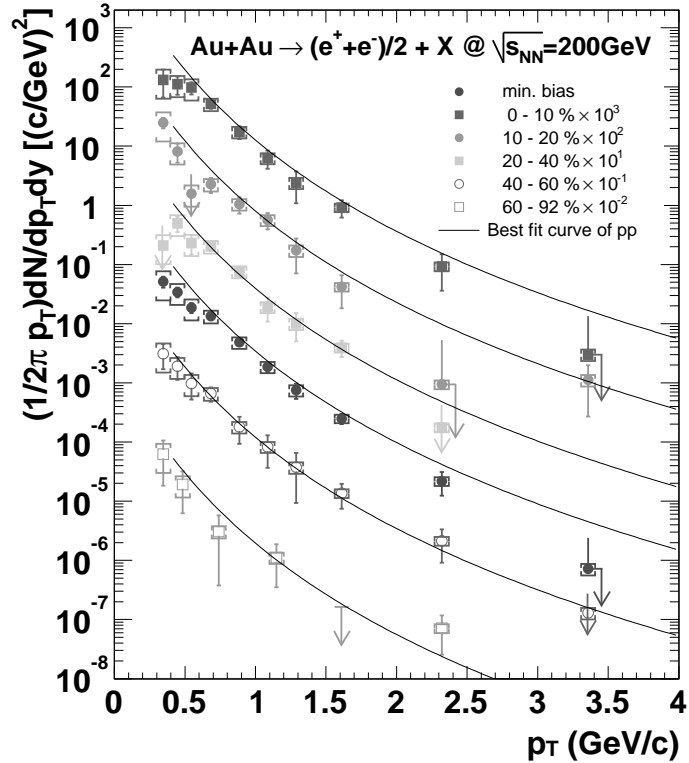}
\caption{Invariant spectra of electrons from non-photonic sources
measured in Au-Au collisions at $\sqrt{s_{NN}}$ = 200 GeV in various
centrality classes (from Ref.~\cite{phenix_au200}).}
\label{fig3}
\end{center}
\end{figure}

\begin{figure}
\begin{center}
\includegraphics[width=0.7\textwidth]{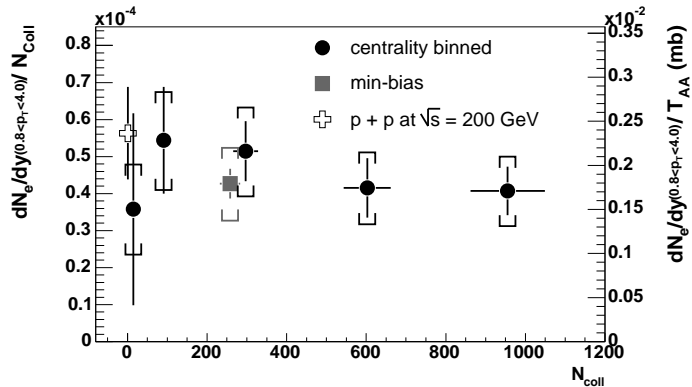}
\caption{Integrated non-photonic electron yield per binary collision
$N_{coll}$ in the range $0.8 < p_T < 4.0$~GeV/c as function of $N_{coll}$ 
in Au-Au collisions at $\sqrt{s_{NN}}$ = 200 GeV (from 
Ref.~\cite{phenix_au200}).}
\label{fig4}
\end{center}
\end{figure}

Fully corrected spectra of non-photonic electrons from Au-Au collisions at
$\sqrt{s_{NN}}$ = 200 GeV are shown in Fig.~\ref{fig3} for five centrality
classes \cite{phenix_au200}. As for the d-Au analysis, the converter 
method was used employing data samples of 2.2M converter-in and 2.5M 
converter-out minimum bias events. The centrality dependence of heavy flavor
production can be addressed by calculating the integrated non-photonic
electron yield and fitting it to the function $AN^{\alpha}_{coll}$.
In the absence of medium effects, one would expect heavy flavor production
to scale linearly with $N_{coll}$, {\it i.e.} $\alpha$ = 1.
This consistency check with the binary scaling hypothesis is shown in
Fig.~\ref{fig4}.
We find $\alpha$ = 0.938 $\pm$ 0.075 (stat.) $\pm$ 0.018 (sys.), which 
indicates that the total yield of non-photonic electrons in the considered
$p_T$ range, and therefore the total initial charm yield, is consistent 
with $N_{coll}$ scaling.
The statistics of the non-photonic electron spectra above $p_T \approx 
2.5$~GeV/c are not sufficient to address the important issue of potential 
medium modifications of the phase space distibution of heavy flavor in 
Au-Au collisions at RHIC.
A currently ongoing cocktail analysis of the full Run-2 Au-Au data sample
might provide first insight into this issue, but a detailed investigation 
will require the analysis of the high statistics Au-Au data set that was 
recorded in Run-4.
 
\begin{figure}
\begin{center}
\includegraphics[width=0.6\textwidth]{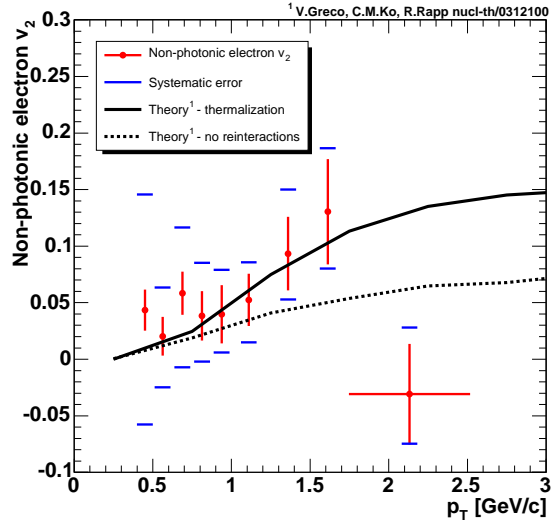}
\caption{Non-photonic electron $v_2$ as function of $p_T$ compared with
model calculations from \cite{greco} employing two different charm flow
scenarios.}
\label{fig5}
\end{center}
\end{figure}

In this context, the measurement of the elliptic flow strength $v_2$ of 
electrons from non-photonic sources is of particular importance since it 
might allow to distinguish different dynamical scenarios from each other.
A preliminary analysis from Run-2 Au-Au data at $\sqrt{s_{NN}}$ = 200 GeV
is shown in Fig.~\ref{fig5}.
The data are compared with two extreme charm flow scenarios.
While the {\it thermalization} scenario assumes that the charm quarks fully
thermalize and participate in hydrodynamic flow on the parton level, the
{\it no-reinteractions} scenario assumes that charm quarks follow unmodified,
pQCD like dynamics and aquire a non-zero $v_2$ only by coalescence with a
flowing light quark \cite{greco}.
Only the analysis of the Run-4 Au-Au data sample will provide enough statistics
to distinguish between these two scenarios.

\section{Summary and Conclusions}

PHENIX has investigated the production of heavy flavor, mainly charm, at 
RHIC in a systematic measurement of electrons from non-photonic sources in
p-p, d-Au, and Au-Au collisions at $\sqrt{s_{NN}}$ = 200 GeV at central
rapidity.
These data imply that the total open charm yield scales with the number of 
binary collisions $N_{coll}$ for all systems and centralities, as expected
for point-like pQCD processes.
Whether the initially produced heavy quarks are then influenced by the
medium in the subsequent dynamical evolution of the nuclear system is
a question of paramount importance.
Detailed answers, however, will have to wait for the analysis of the high
statistics Run-4 data sample where about 1.5B minimum bias Au-Au events
were recorded.
This data set will also allow to broaden the PHENIX open heavy flavor program 
by single muon and lepton pair measurements. Ultimatively, the direct 
reconstruction of particles carrying heavy flavor will become feasible
with a silicon pixel vertex spectrometer which is an integral part of the
PHENIX upgrade program and will facilitate to measurement of the displaced 
vertices of heavy flavor decays \cite{upgrade}.

\end{document}